\documentstyle[12pt,aps,prc,epsfig,preprint]{revtex}
\tightenlines
\newcommand{\beq}{\begin{equation}}
\newcommand{\eeq}{\end{equation}}
\newcommand{\bx}[1]{\mbox{\boldmath $#1$}}
\begin{document}
\begin{titlepage}
\title{\vspace*{10mm}\bf
\Large Study of chirally motivated low-energy $K^-$ optical potentials}
\vspace{6pt}

\author{ A.~Ciepl\'y$^{a,b}$, E.~Friedman$^a$, A.~Gal$^a$,
J.~Mare\v{s}$^b$\\
$^a${\it Racah Institute of Physics, The Hebrew University, Jerusalem 91904,
Israel\\}
$^b${\it Nuclear Physics Institute, 25068 \v{R}e\v{z}, Czech Republic}}

\vspace{4pt}
\maketitle

\begin{abstract}
The $K^-$ optical potential in the nuclear medium is evaluated self
consistently from a free-space $K^-N$ $t$ matrix constructed
within a coupled-channel chiral approach to the low-energy $\bar K N$
data. The chiral-model parameters are fitted to a select subset of
the low-energy data {\it plus} the $K^-$ atomic data throughout the
periodic table. The resulting attractive $K^-$ optical potentials are
relatively `shallow', with central depth of the real part about 55 MeV,
for a fairly reasonable reproduction of the atomic data with
$\chi^2 / N \approx 2.2$. Relatively `deep' attractive potentials
of depth about 180 MeV, which result in other phenomenological
approaches with $\chi^2 / N \approx 1.5$, are ruled out within
chirally motivated models. Different physical data input is required to
distinguish between shallow and deep $K^-$ optical potentials.
The ($K^{-}_{\rm stop},\pi$) reaction could provide such a test,
with exclusive rates differing by over a factor of three for the 
two classes of potentials.
Finally, forward ($K^-,p$) differential cross sections for the
production of relatively narrow deeply bound $K^-$ {\it nuclear}
states are evaluated for deep $K^-$ optical potentials,
yielding values considerably lower than those estimated before.
\newline \newline
$PACS$: 21.65.+f; 24.10.Ht; 36.10.Gv
\newline
$K^-$ nuclear states; $(K^{-}_{\rm stop},\pi)$, 
$(\pi^+,K^+)$ and $(K^-,p)$ reactions
\newline \newline
Corresponding author: E. Friedman, \newline
Tel: +972 2 658 4667,
FAX: +972 2 658 6347, \newline
E mail: elifried@vms.huji.ac.il
\end{abstract}
\centerline{\today}
\end{titlepage}

\section{Introduction}
\label{sec:int}

There is a considerable interest in exploring the behavior
of antikaons in nuclei and in dense nuclear matter \cite{RHK01}.
The issues at stake concern the possibility of witnessing precursor
phenomena, or even the onset, of kaon condensation in dense nuclear
matter, as realized in heavy-ion collisions (see Ref. \cite{CBr99}
and references cited therein) or in neutron stars
(see Ref.~\cite{HHJ00} for a recent review).
At present, the main evidence for a strong
in-medium modification of the $\bar KN$ interaction is due to the
enhanced production of $K^-$ mesons observed in subthreshold and
near-threshold heavy-ion collisions in the KaoS experiments at GSI
\cite{LSB99,LBD00,MBD00}.
The extrapolation of the $\bar K$ nucleus interaction from standard
nuclear density and zero temperature to the higher densities and
temperatures which are relevant for the above phenomena is of course
model dependent. Nonetheless, direct information on the 
$\bar K$ nucleus interaction at standard nuclear density and at zero
temperature is extremely valuable and studies along these lines
are therefore desirable. Unfortunately, even in this
regime the situation is not clear cut at present.

Systematic data bearing on the $\bar K$ nucleus interaction near
threshold are almost exclusively limited to the strong-interaction
shifts and widths of $K^-$ atomic levels throughout the periodic table
\cite{BFG97}. Although the `atomic' $K^-$ meson probes mostly regions
of low nuclear density, the wealth of these data largely compensates
in producing an effective constraint on the extrapolation to higher
densities. The calculations existing todate for the $\bar K$ nucleus
interaction at threshold essentially give two different predictions
for the depth of the $K^-$~nucleus potential at nuclear matter density.
The phenomenological density dependent (DD) optical potential fits
to the kaonic atom data \cite{FGB93,FGB94}, and the relativistic mean
field (RMF) model calculations by Friedman et al. \cite{FGM99},
describe the data quantitatively very well, producing a deeply
attractive potential ($-$Re $V_{\rm opt}(\rho_0) \approx 150-200$~MeV).
In contrast, chirally inspired models of the $\bar KN$ interaction, 
due to Weise and collaborators \cite{KSW95,WKW96} and due to Oset and 
Ramos \cite{ORa98}, give very good fits to the low energy scattering and
reaction data in the strangeness $S=-1$ meson-baryon coupled channel
sector and to the $K^-p$ capture from rest branching ratios, but generally
do not describe well the atomic data. Recently,
following a suggestion by Lutz \cite{Lut98}, the $K^-$ optical potential
has been evaluated {\it self consistently} within such models
\cite{ROs00,SKE00}, yielding qualitatively reasonable fits to kaonic atoms
\cite{HOT00}. Self consistency means that the outcome $K^-$ optical
potential should be accounted for in the in-medium $K^-$ propagator that
generates it within the appropriate scattering integral equation.
These calculations predict a relatively shallow attractive potential
($-$Re~$V_{\rm opt}(\rho_0) \approx 40-60$~MeV).
Very recently, Baca et al. \cite{BGN00} improved significantly the fit
to the atomic data by adding to the self consistent microscopic 
optical potential of Ramos and Oset \cite{ROs00} a phenomenological 
$s$-wave `$t\rho$' term 
of a moderate size (about 30\% increase of the real part
attraction, but 50\% decrease of the imaginary part absorption).
However, this improvement was achieved at the cost of losing the
direct connection of the optical potential to the 
chirally inspired microscopic model of the $\bar KN$ interaction. 

In the present paper, we aim at preserving the above mentioned connection
by showing that reasonable parameters of the chirally motivated 
microscopic model of the $\bar KN$ interaction can be found
such that the low-energy $\bar KN$ data plus the $K^-$ atomic data
are fitted {\it simultaneously}. This is accomplished in Section 
\ref{sec:opt}, where the $K^-$ optical potential providing
the atomic fit is constructed self consistently from the in-medium
$\bar KN$ $t$ matrix. This in-medium quantity reduces to the
free-space $t$ matrix of the chirally motivated model as the density goes
to zero. The calculations presented in Section \ref{sec:opt} 
lead to a relatively shallow potential, 
Re $V_{\rm opt}(\rho_0) \approx -55$ MeV.
This depth is quite similar to that found in Ref. \cite{BGN00}, but the
imaginary part derived in the present work differs substantially from theirs.
The quality of the $K^-$ atomic fit provided by our optical
potential is superior.

The vast difference in potential depths
between the phenomenological and the more microscopic approaches inevitably
brings up the question whether at all, and to what extent, the $K^-$ atom
data contain unambiguous information on the $\bar K$ nuclear potential
at nuclear-matter densities. Our experience leads us to believe that
the behavior of the optical potential at densities about $\rho_0$ is
primarily determined by the functional form, or by the theoretical
model used for the extrapolation from the low-density region which is
more directly connected  to the $K^-$ atom data.
Furthermore, a related question is whether there exists any experimental
procedure for deciding which extrapolation is physically valid.
This question is addressed in the present work by showing that the outcome
of $K^-$ initiated reactions at low energy is sensitive to the wavefunction
of the $K^-$ meson inside the nucleus, where different optical potentials
produce noticeably different wavefunctions. As an example, we discuss
in Section \ref{sec:capture} 
the $(K^{-}_{\rm stop}, \pi)$ reaction into specific hypernuclear states,
demonstrating that the production cross sections are quite sensitive
to the $K^-$ nucleus optical potential.

Several authors \cite{Kis99,AYa99} have drawn attention to $K^-$
{\it nuclear} states bound by a very deep potential, similar to the
best-fit DD potential of Friedman et al. \cite{FGB93,FGB94}. The deepest
states, bound by over 100 MeV, are blocked from decaying by the
two-body mode $\bar KN \to \pi \Sigma$ and their width could then be
reduced to about 10 MeV. If a deeply bound state of this kind is 
identified experimentally, it will provide evidence for a deep 
$K^-$ optical potential. In Section \ref{sec:DD} we evaluate the 
$(K^-, p)$ forward cross sections for production 
of such deeply bound $K^-$~{\it nuclear} states. Our evaluation
yields considerably lower values 
than the estimates made by Kishimoto \cite{Kis99} for this reaction 
at 1 GeV/c.

\section{Optical potentials for kaonic atoms}
\label{sec:opt}

\subsection{Empirical potentials based on $\bar{K}N$ amplitudes}
\label{sec:a(rho)}

$K^-$-nucleus `microscopic' optical potentials, i.e. potentials constructed
from $K^-p$ and $K^-n$ scattering amplitudes as function of the local density,
have recently been discussed by Baca, Garc\'{\i}a-Recio and Nieves in connection
with deeply bound kaonic atom states \cite{BGN00}. In this model the amplitudes 
are calculated self consistently within a chiral approach \cite{ROs00} 
and the optical potential is then constructed by multiplying these amplitudes 
by the corresponding proton and neutron densities. It was shown \cite{BGN00}
that such potentials yield marginally acceptable fits to kaonic atom data
and that phenomenological modifications of the potential lead to much improved
fits to the data. We begin the present section by adopting a similar approach.

The interaction of a $K^-$ meson with the nucleus in a kaonic atom is described
by the Klein-Gordon equation of the form:

\begin{equation}
\label{eq:KG1}
\left[ \nabla^2 - 2{\mu}(B+V_{\rm opt} + V_{\rm c}) + 
(V_{\rm c}+B)^2\right] \psi = 0~~ ~~(\hbar = c = 1) \;\;\; ,
\end{equation}
where $B$ is the complex binding energy and $V_{\rm c}$ is the finite-size 
Coulomb interaction of the hadron with the nucleus, including
vacuum-polarization terms. Equation (\ref{eq:KG1}) assumes
that the optical potential $V_{{\rm opt}}$ behaves
as a Lorentz scalar. It is given by a `$t(\rho)\rho$' form:

\begin{equation}
\label{eq:Vopt}
2\mu V_{{\rm opt}}(r) =
 -{4\pi}(1+\frac{\mu}{M})[a_{K^-p}(\rho) \rho_p (r)
+a_{K^-n}(\rho) \rho_n (r)]
\end{equation}
where $M$ is the nucleon mass, $\mu$ is the $K^-$-nucleus reduced mass,
$a_{K^-p}$ and $a_{K^-n}$ are the $K^-p$ and $K^-n$ threshold scattering 
amplitudes evaluated at a nuclear matter density $\rho = \rho_p + \rho_n$, and
$\rho_p (r)$ and $\rho_n (r)$ are the proton and neutron density distributions. 
The degree of success of any given potential in reproducing the experimental
results for kaonic atoms is represented by the values of $\chi ^2$,
defined in the
usual way. However, comparing the values of $\chi ^2$ achieved in the present
work (see below) with those of Baca et al. \cite{BGN00}, one notes that our
values are higher. The reason for that is most likely the fact that we use
the experimental values for the yields of `upper' levels \cite{FGB94}
whereas Baca et al. seem to use the {\it derived} widths. It is easy to see
that when the experimental accuracy is not  high, the two definitions
of the corresponding $\chi ^2$ may lead to very different $\chi ^2$ values.
The use of the measured quantities in the definition of the $\chi ^2$
function rather than derived quantities is the more appropriate way. 

Applying the amplitudes of Ramos and Oset \cite{ROs00} without any free
parameter leads to $\chi ^2 = 300$ or $\chi ^2/N$ $ (\chi ^2$ per point)
of 4.62. This value should be compared with $\chi ^2/N=1.49$ obtained for 
a phenomenological potential \cite{FGB94} satisfying the low-density theorem 
constraint. The depths of the potentials at the center 
of a typical nucleus such as Ni are 44 and 54 MeV for the real and imaginary
parts, respectively, compared to 187 and 71 MeV
respectively for the phenomenological potential.
In order to improve the fit to the data we followed Baca et al. \cite{BGN00}
and added a phenomenological  `$t\rho$' term to the potential. Another
possibility is to apply some scaling factors to the input amplitudes.
In order to preserve the predictions of the chirally motivated model in the 
isospin zero sector where the strongly coupled channels produce the 
$\Lambda$(1405) resonance, we have applied scaling
factors $s_{\rm R}$ and $s_{\rm I}$ only to the isospin 1 combination of the
$K^-p$ and $K^-n$ amplitudes.
More specifically, we have

\begin{equation}
  a_{I=0} = 2a_{K^-p}-a_{K^-n},\quad \quad  a_{I=1} = a_{K^-n} ,
\end{equation}
and then we set
\begin{equation}
  a'_{I=1} = s_{\rm R} \:{\rm Re}\, a_{K^-n}
+ {\rm i} s_{\rm I} \:{\rm Im}\, a_{K^-n}
\end{equation}
and derive the modified scattering amplitudes as
\begin{equation}
a'_{K^-n} = a'_{I=1},\quad \quad a'_{K^-p} = \frac{1}{2} (a_{I=0} + a'_{I=1}).
\label{eq:scaling}
\end{equation}

Table \ref{tab:ROS} summarizes the results
obtained using the input amplitudes, with or without adding 
a `$t\rho$' term and with or without introducing scaling factors $s$. Also
included in the table are similar results for the amplitudes of
Schaffner-Bielich et al. \cite{SKE00}. Rather poor fits
to the data are obtained when the potentials are constructed from 
the input amplitudes. The fits improve significanly when 
an empirical `$t\rho$' potential is added, but  the resulting total 
potential is not unique, namely, potentials giving similar values of 
$\chi ^{2}$ may differ by a factor of 2-3 in the nuclear interior. 
For the Ramos and Oset \cite{ROs00} input amplitudes, we essentially 
confirm the depths originally found in Ref. \cite{BGN00} upon applying 
this procedure. Further improvements in the fits are found when 
scaling factors, as defined by Eq. (\ref{eq:scaling}), are included. 
The values of  $\chi ^2/N$ may
approach the best value obtained with phenomenological potentials
\cite{BFG97} but the required scaling factors are likely to be unphysical.

\subsection{$\bar{K}N$ scattering amplitudes in the nuclear medium}
\label{sec:KN}

Here we follow the chirally motivated model of Ref.~\cite{KSW95}
for $\bar KN$ scattering and reactions near threshold.
The $\Lambda$(1405) subthreshold resonance is generated in this model
dynamically by solving coupled Lippmann-Schwinger equations for the
$t_{ij}$ elements of the $t$ matrix in terms of the input
chiral potentials $v_{ij}$. The six coupled meson-baryon channels
included in this model are $K^-p$, $\bar{K}^{0}n$, $\pi^{0}\Lambda$,
$\pi^{+}\Sigma^{-}$, $\pi^{0}\Sigma^{0}$ and $\pi^{-}\Sigma^{+}$.
The use of a nonrelativistic formalism is justified for energies
close to the $\bar{K}N$ threshold in accordance with the
aims of the present paper.

Applying the coupled-channel model to the calculation of the {\it in-medium}
$t(\rho)$ matrix, the density dependence of the $\bar K$ optical potential
$t_{\bar KN}(\rho)\rho$ can be traced to the propagation of the
$\Lambda$(1405) resonance in the nuclear medium. Waas et al. \cite{WKW96},
following Koch \cite{Koc94}, demonstrated that Pauli blocking of the
intermediate nucleon is primarily responsible for the transition from
a repulsive $t_{\bar KN}(\rho = 0)$, consistently with the
subthreshold $\Lambda$(1405), to an attractive $t_{\bar KN}(\rho)$ in the
nuclear interior, as required by the $K^{-}$-atomic data.
Later, Lutz \cite{Lut98} stressed the importance of including the kaon
self energy within a self consistent calculation,
and Ramos and Oset \cite{ROs00} provided a comprehensive calculation
of this kind including also pion, nucleon and hyperon self energies.

We extend the calculations of Refs. \cite{KSW95,WKW96} by including
$\bar K$ and $N$ self energies. In this approach, the chirally motivated
coupled-channel potentials are taken in a separable form
\beq
v_{ij}(k,k')=\frac{C_{ij}}{f_{\pi}^2}\beta_i\beta_j
g_{i}(k^{2})g_{j}(k'^{2}), \;\;\;\;
g_{j}(k)=\frac{1}{1+(k/ \alpha_{j})^2} \; ,
\label{eq:vij}
\eeq
where the momenta $k$ and $k'$ refer to the meson-baryon
c.m. system in the $i$ and $j$ channels, respectively, and the relativistic flux
normalization factors $\beta_i$ are defined by
\beq
\beta_i = \sqrt{\frac{1}{2\omega_i}\frac{M_i}{E_i}} \; ,
\eeq
with $\omega_i$, $M_i$ and $E_i$ denoting the meson energy, the baryon mass
and energy in the c.m. system of channel $i$. The coupling
matrix $C_{ij}$ is determined by chiral SU(3) symmetry and includes terms
up to second order in the meson c.m. kinetic energies
(see Ref.~\cite{KSW95} for more details). Finally, the parameter $f_{\pi}=94.5$ MeV
represents the pseudoscalar meson decay constant,
and the inverse range parameters $\alpha_{i}$ were
fitted to the low energy $\bar{K}N$ data in Ref.~\cite{KSW95}. Their values are
\begin{eqnarray}
\alpha_{K^-p} & = & \alpha_{\bar K^0n}=757.8 \;{\rm MeV}\;, \nonumber \\
\alpha_{\pi^0\Lambda} & = & 300 \;{\rm MeV}\;, \label{eq:alpha} \\
\alpha_{\pi^+\Sigma^-} & = & \alpha_{\pi^0\Sigma^0}=
\alpha_{\pi^-\Sigma^+}=448.1 \;{\rm MeV}\;. \nonumber
\end{eqnarray}

The elementary amplitudes corresponding to the potentials (\ref{eq:vij})
are of the form 
\beq
f_{ij}(k,k';E)=-\frac{1}{4\pi f_{\pi}^{2}} \sqrt{\frac{M_{i}M_{j}}{E^{2}}}
g_{i}(k^{2}) g_{j}(k'^{2})
\left[ (1 - C \cdot G(E))^{-1} \cdot C \right]_{ij}\;\; ,
\eeq
where the meson-baryon propagator $G(E)$ is diagonal in the channel indices
$i$ and $j$ and is given by
\beq
G_{i}(E;\rho)=\frac{1}{f_{\pi}^2}\frac{M_i}{E}\int_{\Omega_{i}(\rho)}
\frac{d^{3}p}{(2\pi)^{3}}\frac{g_{i}^{2}(p^{2})}{k_{i}^{2}-p^{2}
-\Pi_{i}(\omega_{i},E_{i},\bx{p};\rho)+{\rm i}0}\;\; .
\label{eq:Green}
\eeq
Here the integration domain $\Omega_{i}(\rho)$ is limited by the Pauli
principle in the $\bar{K}N$ channels, $\rho$ denotes the nuclear density and
$k_{i}$ is the on-shell c.m. momentum in channel $i$, such that
$\omega_{i}^{2}=m_{i}^{2}+k_{i}^{2}, E_{i}^{2}=M_{i}^{2}+k_{i}^{2}$,
and $E=E_{i}+\omega_{i}$ is the total energy. In the denominator of the Green's 
function (\ref{eq:Green}) we have included the meson plus baryon self energy
$\Pi_{i}$. For simplicity, we neglect the self energy corrections
in the pion-hyperon channels, expecting their effect on the $\bar{K}N$ channels
to be secondary to the primary effect of including the $\bar K$ and nucleon
self energies \cite{SKE00}. However, the pion self-energy effect was
found nonnegligible in Ref.~\cite{ROs00} and this point deserves further study.

The self energy term $\Pi$ in the c.m. $\bar KN$ channels, close to threshold,
is expressed in terms of the more familiar self
energies of the antikaon and the nucleon as follows:
\beq
\label{eq:field}
\Pi=\frac{\mu_{KN}}{m_K}\Pi_{\bar K}+
\frac{\mu_{KN}}{M}\Pi_N\; ,
\eeq
where $\mu_{KN}$ is the $\bar KN$ reduced mass,
$\Pi_{\bar K}=2m_{K}V_{\rm opt}^{\bar K}$ in terms of the $\bar K$
optical potential given in Eq. (\ref{eq:Vopt}),
and $\Pi_N=2MV_{\rm opt}^N$ in terms of the nucleon optical potential
which in the present work was taken in the form
\beq
\label{eq:vnn}
V_{\rm opt}^N=V_{0}\frac{\rho}{\rho_{0}}\;\; ,
\eeq
where $\rho_0 = 0.17$~fm$^{-3}$. For the nucleon optical
potential we used $V_{0} = (-60 - {\rm i} 10)$~MeV. The real part is
consistent with mean-field potentials used in nuclear structure
calculations. The strength of the imaginary part is adopted from
proton-nucleus scattering analysis \cite{ACMS81}.

It should be noted that when the meson and baryon self energies are
turned off, our model reduces to the in-medium model presented
in Ref.~\cite{WKW96}. It is also obvious that the same applies
for zero density. This means that we are able to reproduce the
free-space threshold branching ratios and cross sections fitted in
Ref.~\cite{KSW95}. Below we  show that a satisfactory description
of these free-space low-energy data can also be
achieved in a simultaneous fit of the model parameters to these data and
to the $K^{-}$-atomic data.

\subsection{Fits to kaonic atoms and $K^-p$ data}
\label{sec:fits}

The chiral model presented above was applied in $\chi^{2}$ fits to
$K^{-}$-atomic data and to representative low energy $K^-p$ data.
The latter consist of the three accurately measured threshold
branching ratios \cite{Mar81}
\beq
\gamma  =  \frac{\Gamma(K^{-}p \rightarrow \pi^{+}\Sigma^{-})}
             {\Gamma(K^{-}p \rightarrow \pi^{-}\Sigma^{+})} \;, \; \;
R_{\rm c}   =  \frac{\Gamma(K^{-}p \rightarrow {\rm charged})}
             {\Gamma(K^{-}p \rightarrow {\rm all})} \;, \; \;
R_{\rm n}  = \frac{\Gamma(K^{-}p \rightarrow \pi^{0}\Lambda)}
             {\Gamma(K^{-}p \rightarrow {\rm neutral})} \;, \; \;
\label{eq:ratios}
\eeq
plus four $K^-p$ - initiated cross sections at 110 MeV/c to the channels
other than the $\pi^{0}\Lambda$ and $\pi^{0}\Sigma^0$ channels for which
the quality of data is inferior. The adopted experimental values of the
cross sections \cite{Mar81,Cib82,Eva83} 
\beq
\sigma(K^{-}p \rightarrow K^{-}p) \;,  \; \;
\sigma(K^{-}p \rightarrow \bar{K}^{0}n) \;, \; \;
\sigma(K^{-}p \rightarrow \pi^{+}\Sigma^{-}) \;, \; \;
\sigma(K^{-}p \rightarrow \pi^{-}\Sigma^{+}) \;, \; \;
\label{eq:xsecs} 
\eeq  
as well as the measured branching ratios $\gamma$, $R_{\rm c}$ and 
$R_{\rm n}$ are listed in Table \ref{tab:ratios}. The values calculated 
for the above seven quantitities are also listed, using $(i)$ the 
original parameterization of Ref. \cite{WKW96} (denoted `no SC') 
which did not account for $\bar K$ and $N$ self energies; 
and $(ii)$ the present model (denoted `V$_{\rm N}$, full') which 
was fitted to both the free-space $K^-p$ data 
(\ref{eq:ratios},\ref{eq:xsecs}) as well as to the $K^-$ atomic data, 
including self consistently $\bar K$ and $N$ self energies. 
Table \ref{tab:ratios} demonstrates that extending the fit to 
include the atomic data does not spoil the good agreement with 
the experimental low energy $K^-p$ data. 

Table \ref{tab:coupl} summarizes the results obtained within the
present approach. The upper two rows involve no fits to the atomic data;
the $\bar K$ self energy is not included in the in-medium calculation
pertaining to the first row (`no SC'), whereas this self energy is
included self consistently in the in-medium amplitudes of the calculation
of the second row (`SC').
The nucleon self energy is excluded in both calculations. The $\chi^{2}$
value for the seven $K^-p$ data points is of course the same in both
calculations which, however, differ markedly in the $\chi^{2}_{\rm atom}$
values with respect to the 65 atomic data points. The calculated
$K^-$ optical potential is made considerably shallower and the description
of the atomic data improves upon requiring self consistency, in agreement
with Refs. \cite{HOT00,BGN00}. The remaining four 
rows are for various fits to the 65 $K^-$ atomic data points, 
some excluding (`atoms') and some including (`full') the 7 free-space $K^-p$ 
data points in the fit, all with the $\bar K$ self energy included self
consistently in the in-medium calculation. The last two rows (`$V_{\rm N}$')
are for the additional inclusion of the nucleon self-energy optical
potential of Eq. (\ref{eq:vnn}). The table also specifies the resulting scaling 
factors which multiply the pseudoscalar meson decay constant 
$f_{\pi}$ = 94.5 MeV and the inverse 
range parameters $\alpha_i$ of Eq. (\ref{eq:alpha}). 
It is seen that the modification of the free-space model parameters 
is moderate. Also given are the depths of the real and imaginary 
potentials at the center of the Ni nucleus. 
The depth of 55 MeV for the real part (last row in the table) is very close 
to the corresponding depth found in Ref. \cite{BGN00}, but the depth of 60 MeV 
for the imaginary part is considerably larger than that found there. 
It is clear from the table that, with relatively small modifications of
the chiral-model parameters, it is possible to achieve reasonably good fits
to both atomic and free-space data. The best  $\chi ^2_{\rm atom}$
means $\chi ^2_{\rm atom}/N$ of 2.2, a value which is significantly lower than
the value 2.7 corresponding to the hybrid fit of Ref. \cite{BGN00}, yet
considerably higher than the value 1.5 obtained with best-fit phenomenological
potentials \cite{BFG97}. It is therefore again concluded that when a free-space
interaction model is used to fit the $K^-$ atomic data, the quality of the fit
is inferior to that achieved with phenomenological fits which are constrained
only by the low density theorem.

\begin{figure}
\epsfig{file=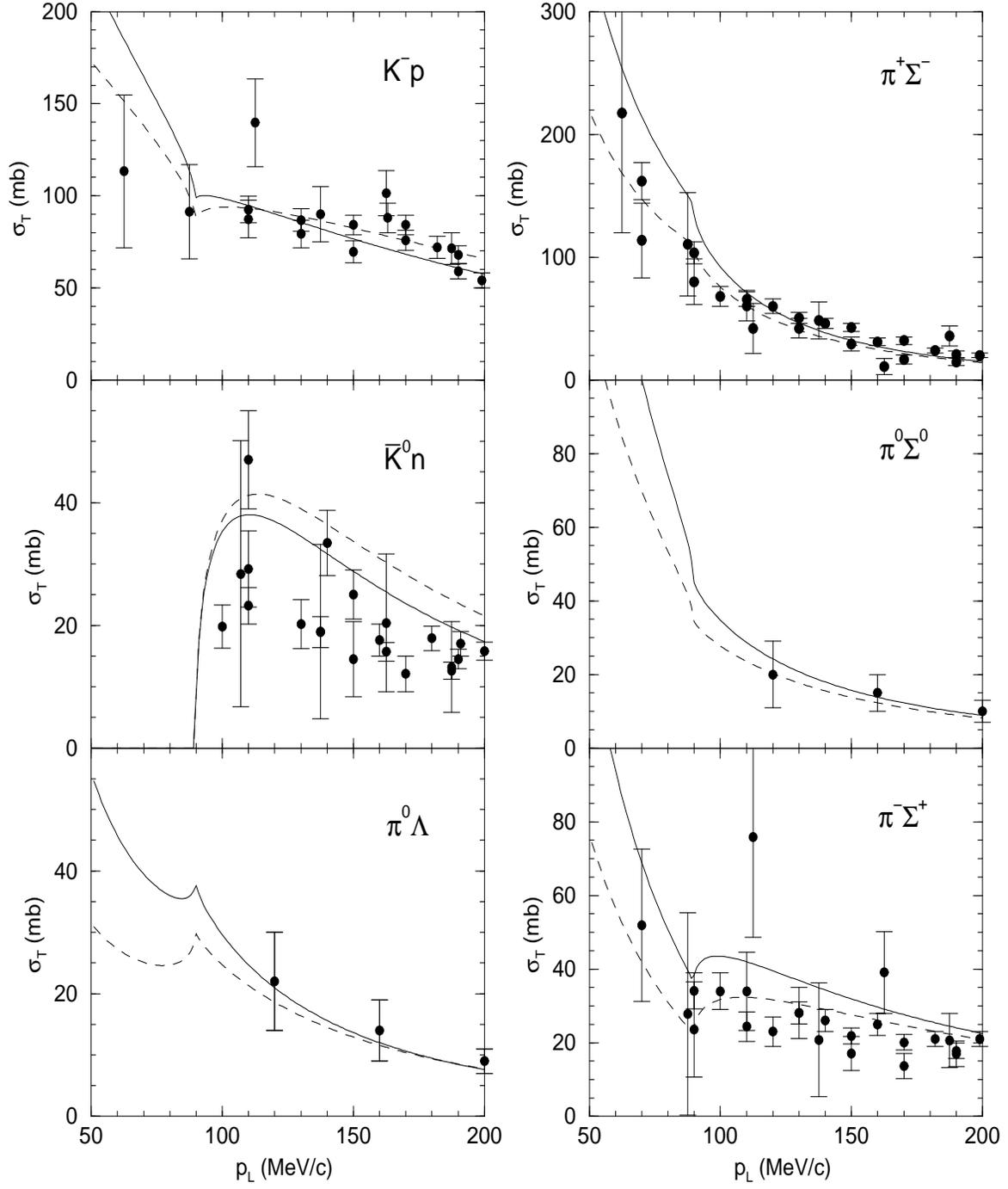, height=185mm,width=155mm}
\vspace*{5mm}
\caption{
Cross sections for $K^{-}p$ scattering and reactions to the channels 
indicated in the figure. Results of the full fit 
(`$V_{\rm N}$, full', see Table \ref{tab:coupl}), displayed 
by solid lines, are compared with the available data (see text). 
Results for the free-space chiral-model parameterization (`no SC') 
are shown for comparison (dashed lines).} 
\label{fig:xs}
\end{figure}

Figure \ref{fig:xs} shows cross sections calculated for
six free-space $K^-p$ initiated reactions in comparison with the data 
(as compiled in Fig. 1 of Ref. \cite{KSW95}). 
The results of our full model (`$V_{\rm N}$, full') and the
free-space calculation
(`no SC') were chosen for illustration. The cross sections calculated for
the other
parameterizations of Table \ref{tab:coupl} are of a similar quality.
It is emphasized that we obtain equally good fits to
the cross sections as well
as to the threshold branching ratios even when {\it only} $K^-$ atomic
data are used in the fit.
Moreover, we find that the inclusion of the nucleon
self-energy optical potential
$V^{N}_{\rm opt}$ plays only a marginal
role in the improvement of the fit to the data.
Of course, the size and form of $V^{N}_{\rm opt}$ influences
considerably the depth of
the resulting $K^-$ nucleus optical potential, as demonstrated in the table.

\begin{figure}
\epsfig{file=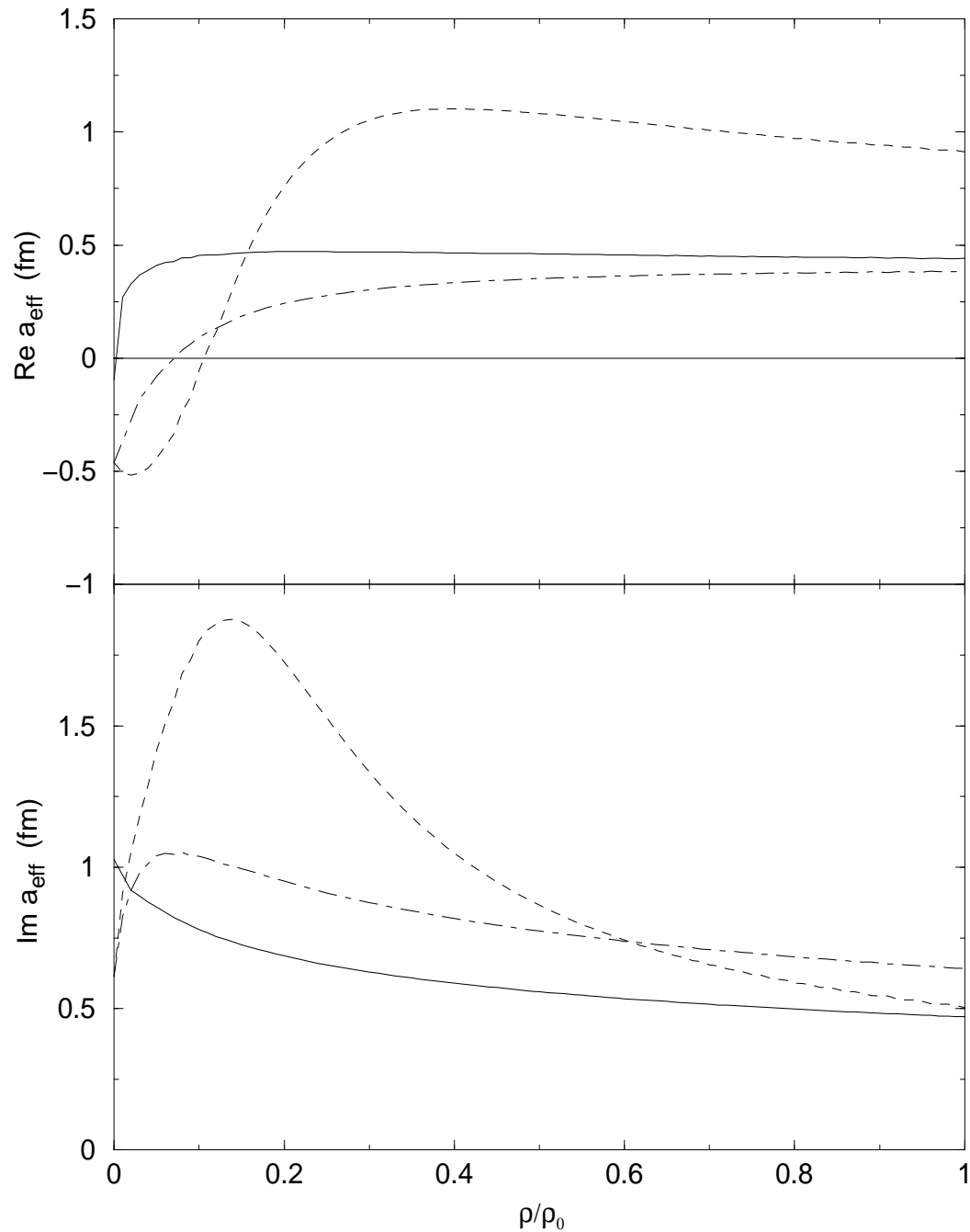, height=180mm,width=140mm}
\vspace*{7mm}
\caption{
Real (top) and imaginary (bottom) parts of the isospin-averaged 
$K^- N$ (effective) threshold scattering amplitude as function of density 
$\rho/\rho_0$, calculated for the `no SC' (dashed line), `SC' 
(dot-dashed line) and `$V_{\rm N}$, full' (solid line) 
chiral-model parameterizations (see Table \ref{tab:coupl}).} 
\label{fig:aef}
\end{figure}

Figure \ref{fig:aef} shows the density dependence of the
isospin averaged (effective) threshold scattering amplitude 
$a_{\rm eff} = (3a_{\rm eff}^{I=1}+a_{\rm eff}^{I=0})/4$ 
for three cases selected from Table \ref{tab:coupl}: $(i)$ no medium 
effects beyond Pauli blocking are included (`no SC', dashed line); 
$(ii)$ the self-consistent calculation including the $\bar K$ self energy
(`SC', dot-dashed line); and $(iii)$ plus including the nucleon self energy
(`$V_{\rm N}$, full', solid line).
The change of the sign of Re $a_{\rm eff}$ from negative to positive
corresponds to the transition from an apparently repulsive free-space 
interaction to an attractive one in the nuclear medium.
In the `no SC' model, in which medium modifications are represented only by
the Pauli blocking effect, the transition occurs at
$\rho \approx 0.1 \rho_0$. When the $K^-$ self energy is taken into account,
this transition occurs at a lower density, $\rho \approx 0.05 \rho_0$, and
the inclusion of $V^{N}_{\rm opt}$ pushes this transition density even further 
down. Nevertheless, the free-space ($\rho = 0$) threshold scattering amplitude 
remains negative even in this case, reflecting the dominance of the 
$\Lambda$(1405) $I=0$ subthreshold resonance. We note that since the 
`no SC' and `SC' parameterizations coincide, the low-density limit of 
$a_{\rm eff}$ for these cases is the same, whereas $a_{\rm eff}(\rho = 0)$ 
for the `$V_{\rm N}$, full' parameterization assumes a different value. 

Taking the $K^-$ self energy into account generally leads to a weaker
density dependence of the threshold scattering amplitude, both for its real
and imaginary parts. This indicates that the $K^-$ optical potential
evaluated within such self consistent models is well approximated 
by a $t_{\rm eff}\rho$ form (where $t_{\rm eff}$ = const.) over a wide 
range of densities. A genuine $\rho$ dependence of $t_{\rm eff}$ 
appears only at very low densities.

\begin{figure}
\epsfig{file=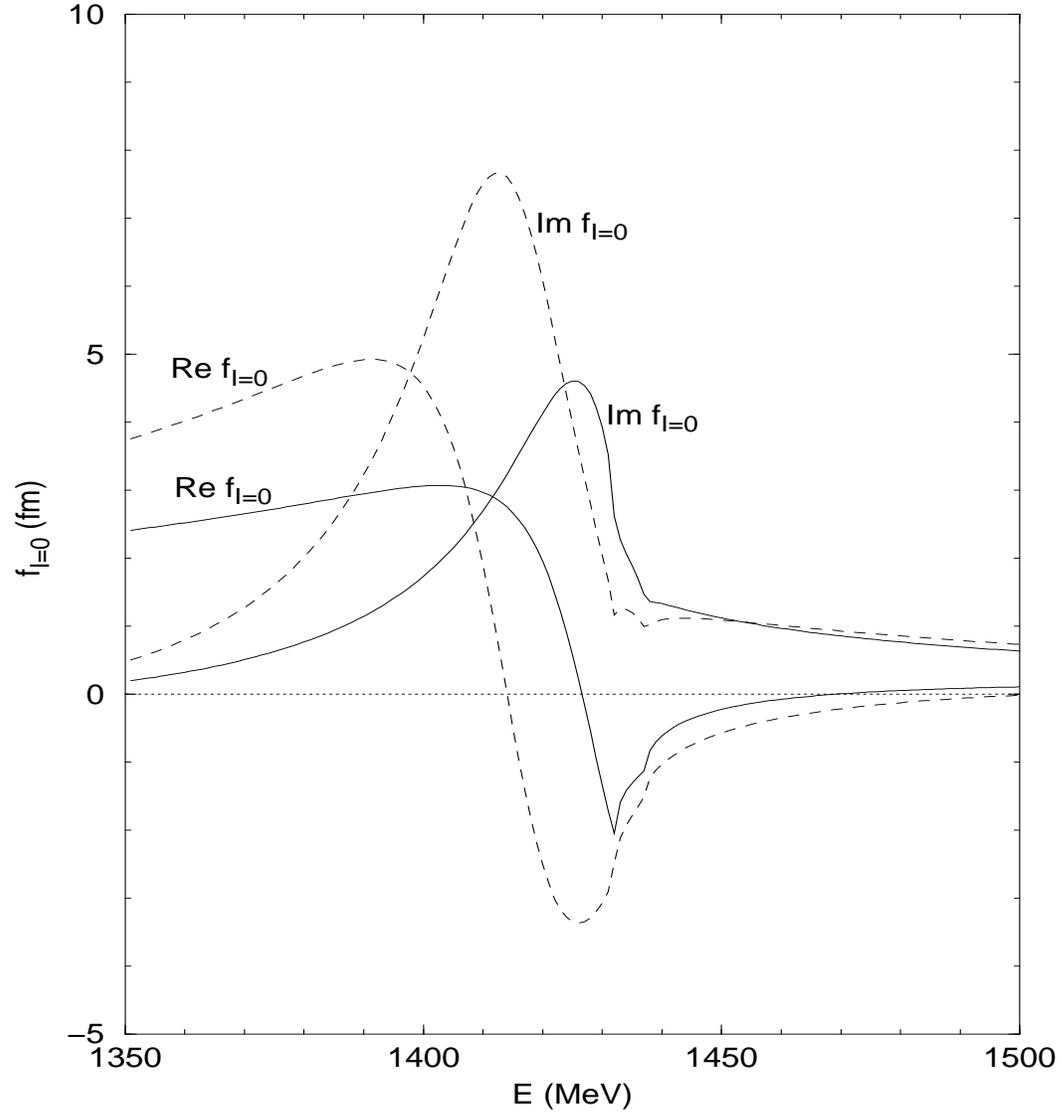, height=150mm,width=140mm}
\vspace*{7mm}
\caption{
The $\bar KN$ free-space scattering amplitude $f_{I=0}$
as function of c.m. energy.
Results are presented for the `no SC' (dashed lines)
and `$V_{\rm N}$, full' (solid lines) chiral-model
parameterizations from Table \ref{tab:coupl}.}
\label{fig:a0}
\end{figure}

The free-space ($\rho = 0$) scattering amplitude $f_{I=0}$, as a function 
of the c.m. energy $E$, is shown in Fig. \ref{fig:a0}.
The calculated amplitude for the `no SC' parameterization (first row in Table 
\ref{tab:coupl}, dashed line here) is compared with that for the 
`$V_{\rm N}$, full' parameterization (last row in Table 
\ref{tab:coupl}, solid line here). The peak of Im $f_{I=0}$ is shifted 
upward by over 10 MeV
when the $K^-$ atom data are included in the fit. These amplitudes, 
for which the real part changes sign at the energy where the imaginary 
part peaks, provide a signature of the $\Lambda$(1405) subthreshold 
resonance. We note that a precise reproduction of the $\Lambda$(1405) 
spectral shape requires a more involved calculation than the $I=0$ 
$\bar K N - \pi \Sigma$ coupled channel calculation reported in the 
present work and, therefore, this spectral shape was not used here 
as a constraint. A new datum that was unavailable to the 
authors of Refs. \cite{KSW95,WKW96} is the $K^-p$ scattering length, 
deduced from the recent measurement of the $2p \rightarrow  1s$ X ray 
in kaonic hydrogen \cite{ITO98} and which we too have not included in the 
fit to the data. Model `$V_{\rm N}$, full' does very well with respect 
to Re $a_{K^-p}$, but does poorly for the imaginary part. 
However, the rapid variation of Im $f_{I=0}$ near threshold should 
make it fairly easy to reproduce it by slightly varying the parameters 
of the present chiral model.
 
For completeness, we also tested the 10 channel chiral model of Refs. 
\cite{ORa98,ROs00}, including also the $\eta \Lambda$,
$\eta \Sigma^{0}$, $K^{+}\Xi^{-}$, and $K^{0}\Xi^{0}$ channels. We treated it
on the same footing as our 6 channel model, namely a separable
form was used for the coupled-channel potentials,
and only $\bar K$ and $N$ self energies were considered. Fitting to the $K^-$
atom data led to $\chi^2$ values comparable to those for the
6 channel model.

\section{Stopped $K^-$ reactions as a test of the $K^-$ 
nucleus optical potential}
\label{sec:capture}

In this section we discuss the possibility of testing the 
$K^-$ optical potential at threshold by 
studying ($K^{-}_{\rm stop},\pi$) reactions to specific $\Lambda$ 
hypernuclear states. The ($K^{-}_{\rm stop},\pi^-$) reaction has been 
used to explore the spectroscopy of $\Lambda$ hypernuclei 
(see Ref. \cite{THO94} for a review). Recently, the use of the 
complementary ($K^{-}_{\rm stop},\pi^0$) reaction was proposed 
\cite{AGS94} and first results on $^{12}$C are forthcoming \cite{CRu01}.
Calculations of $\Lambda$-hypernuclear formation rates for
($K^{-}_{\rm stop},\pi^-$) reactions were presented by several
groups \cite{HLW74,GKl86,BMo86,MYa88} within the framework of the
distorted wave impulse approximation. The present calculation
follows the approach described in Ref.\cite{GKl86}. Our aim is to 
study the sensitivity of the capture rates to the choice of the $K^-$~ 
nucleus optical potential, provided the latter was fitted to the $K^-$ 
atomic data. Here we limit the discussion to the 
following $K^-$ capture-at-rest reactions on $^{12}$C:  
\beq
K^{-}\;+\;^{12}{\rm C}\;\longrightarrow \;\pi^{-}\;+\;_{\Lambda}^{12}{\rm C} 
\;,\;\;\;\;\; K^{-}\;+\;^{12}{\rm C}\;\longrightarrow \;
\pi^{0}\;+\;_{\Lambda}^{12}{\rm B}\;.
\label{eq:reaction}
\eeq

The capture rate per stopped $K^{-}$ from the $^{12}$C initial state 
$i$ to a $\Lambda$-hypernuclear final state $f$, 
$R_{fi}/K^{-}$, is given by \cite{GKl86} 
\beq
R_{fi}/K^{-}= \frac{q_{f}\omega_{f}}{{\bar q}_{f}{\bar \omega}_f}
  \:R(\pi \Lambda)\:
  \frac{\int d\Omega_{\bx{q}_{f}} \langle\,\mid F_{fi}^{\rm DW}(\bx{q}_{f})
 \mid^{2}\,\rangle} 
 {4\pi {\bar \rho}_{N}}\mbox{\hspace*{10mm},}
\label{eq:R/K} 
\eeq 
where the fractions $R(\pi \Lambda)$ are the in-medium branching ratios 
for $K^{-}N \rightarrow \pi \Lambda$ capture at rest, as given in Table 
I of Ref.\cite{GKl86}, and the kinematic factor involving the pion momentum 
$q$ and energy $\omega$ in front of $R(\pi \Lambda)$ 
is due to considering the pion final-state phase space with respect to an 
average closure phase space marked by bars. 
The product of these two factors for capture on $^{12}$C is 
0.074 (for $R(\pi^{-} \Lambda)$) $\times$ 1.456 
(for $q_{f}\omega_{f}/{\bar q}_{f}{\bar \omega}_{f}$), equaling 
0.1077 for ($K^{-}_{\rm stop},\pi^-$) 
and half of that, by charge independence, for ($K^{-}_{\rm stop},\pi^0$). 
The distorted wave (DW) transition amplitude is given by
\beq
F_{fi}^{\rm DW}(\bx{q}_{f})=\int d^{3}r\: \chi^{(-)*}_{\bx{q}_{f}}(\bx{r})
  \:\rho_{fi}(\bx{r})\:\Psi_{nLM}(\bx{r})\mbox{\hspace*{10mm},}
\label{eq:DW}
\eeq
with $\rho_{fi}$ denoting the nuclear to hypernuclear transition density 
matrix element. 
The brackets $\langle \cdots \rangle$ in Eq. (\ref{eq:R/K}) 
stand for averaging on the initial substates and summing over the final ones. 
Finaly, ${\bar \rho}_N$ denotes the effective
nuclear density available to the capture process,
\beq
{\bar \rho}_{N} = \int d^{3}r \:\rho_{N}(r)\: 
\mid \Psi_{nLM}(\bx{r}) \mid^{2} \mbox{\hspace*{10mm},} 
\label{eq:rho}
\eeq
where $\rho_{N}(r)$ is the appropriate nucleon (proton or neutron) 
density and $\Psi_{nLM}(\bx{r})$ is the $K^-$~wavefunction 
in the $nL$ orbit from which the $K^-$ meson is captured. 

The radial wavefunctions for nucleons in $^{12}$C, and for the 
$\Lambda$ hyperon in $_{\Lambda}^{12}$C and $_{\Lambda}^{12}$B, were 
generated by solving the Schr\"{o}dinger equation for a real Woods-Saxon 
potential with a diffusivity parameter $a=0.6$ fm. The depth $V_{0}$ was 
adjusted separately for each baryon orbit to yield the observed binding 
energies. The proton Coulomb potential due to the nuclear core charge 
distribution was also included. The single-particle wavefunctions 
thus obtained were used in the calculation of $\rho_{fi}(\bx{r})$ for 
the DW amplitude (\ref{eq:DW}) and also for constructing $\rho_{N}(r)$ 
in Eq. (\ref{eq:rho}). The value of the radius parameter $r_0$ 
was chosen such that for protons the r.m.s. radius of $\rho_p$ was equal 
to the r.m.s. radius of the charge distribution after unfolding from the 
latter the proton size. As already observed in Ref.\cite{GKl86}, since 
the baryonic radial wavefunctions that enter the capture rate calculation 
are nodeless and real, the resulting calculated rates (\ref{eq:R/K}) 
are considerably less sensitive to variations in these wavefunctions 
than to similar model variations of the pion and $K^-$ wavefunctions.

To generate the pion DW $\chi^{(-)}_{\bx{q}_{f}}$ in Eq. (\ref{eq:DW}) 
we used the measured pion elastic scattering angular distribution 
for $\theta_{\rm cm} < 90^{\circ}$ at 162 MeV on $^{12}$C \cite{PIF77} 
in order to fit the standard pion-nucleus optical potential 
\beq
2\mu_{\pi}V_{\rm opt}^{\pi} = 4\pi[-(1+\frac{\mu_{\pi}}{M})b_{0}\rho(r)
  +(1+\frac{\mu_{\pi}}{M})^{-1}c_{0}{\bf {\nabla}}\rho(r)\cdot{\bf {\nabla}}] 
\label{pipot1}
\eeq 
for which $\chi$ solves the Klein Gordon equation. 
We have managed to improve considerably the fits of Ref. \cite{GKl86}, 
getting as low as $\chi^{2}/N = 3.2$ for the following values of parameters: 
\beq 
b_0 = (-0.24+{\rm i} \; 0.18) \; m_{\pi}^{-1} \;, \;\;\; 
c_0 = ( 0.20+{\rm i} \; 0.31) \; m_{\pi}^{-3} \; . 
\label{pipot2} 
\eeq 

In addition to the chirally motivated $K^-$ optical potential 
discussed in the previous section, 
several $K^-$ initial-state wavefunctions were generated 
using the density-dependent form of the $K^{-}$ optical potential
\cite{FGB94}
\beq
2\mu V_{\rm opt}=-4\pi (1+\frac{\mu}{M})
  \left[ a+B\left(\frac{\rho(r)}{\rho(0)} \right)^{\!\gamma}\,\right] 
  \rho(r) \;\; ,
\label{eq:DD}
\eeq
where the notation follows the `$t(\rho)\rho$' form, 
Eq. (\ref{eq:Vopt}). For these optical potentials, the energy 
shifts and widths of the $2p$ and $3d$ levels in kaonic $^{12}$C, 
and the corresponding wavefunction $\Psi_{nLM}(\bx{r})$, were then 
obtained by solving the Klein-Gordon equation 
(\ref{eq:KG1}). Four different optical potentials, ordered according 
to their central depth, are listed in Table \ref{tab:C1}. 
The `chiral' potential 
corresponds to the relatively shallow potential of the present work. 
The deep potential `DD' was obtained by Friedman et al. \cite{FGB94} 
fitting the parameters $B$ and $\gamma$, with the value of $a$ held 
fixed at the empirical $K^{-}N$ scattering length, so that the 
potential (\ref{eq:DD}) satisfies the low-density limit. The potential 
`$t_{\rm eff}$' was obtained for $B=0$ as the best-fit 
$t_{\rm eff}\rho$ solution for the standard density independent version 
of $V_{\rm opt}$. Since the above potentials were obtained 
within {\it global} fits to the available $K^{-}$-atomic data, 
they do not necessarily 
reproduce precisely the experimental data on kaonic $^{12}$C \cite{BAC72}, 
and for this reason we added in Table \ref{tab:C1} another potential 
(`$\tilde{t}_{\rm eff}$') designed primarily to fit these latter data. 

The calculated capture rates per $K^-$, Eq. (\ref{eq:R/K}), are shown in 
Table \ref{tab:C2} for the production of the $1^-$ hypernuclear ground 
states off $^{12}$C, assuming atomic capture fractions $f_{p}=0.23$ and 
$f_{d}=0.77$ according to a cascade calculation by Batty \cite{BAT95} 
which fits the absolute and relative X-ray intensities for $^{12}$C. 
It is clear that the deeper the $K^-$ optical potential is, 
the lower the calculated rate becomes. This pattern is caused by the 
strong-interaction bound $d$ state generated by all but the `chiral' 
potentials, as may be recognized by the repulsive shift (see Table 
\ref{tab:C1}) they impose on the $3d$ atomic state. By orthogonality, 
the atomic wavefunction acquires then {\it extra} nodes within 
the nucleus, thus causing substantial cancelations in the DW amplitude 
(\ref{eq:DW}). 
This effect was extensively studied in Ref. \cite{GKl86} for $K^-$ capture 
from atomic $p$ states in $^{12}$C, but for the dominant capture from 
atomic $d$ states there was no similar effect since the relatively shallow 
potentials considered there (similar to the present `chiral' potential) 
did not produce strong-interaction bound $d$ states. All the calculated 
rates shown in the table are lower than the measured values. Tamura 
et al. \cite{THO94} report a rate of $(0.98 \pm 0.12) \times 10^{-3}$ per 
stopped $K^-$ for the production of $_{\Lambda}^{12}$C. This value is still 
four times larger than our highest calculated value, that for the `chiral' 
potential. The largest uncertainty in our calculation is due to the pion 
distortion effects which we estimate as about 10\%, so the discrepancy 
between experiment and calculation cannot be resolved at present. 
Therefore, although the $K^-$ capture from rest reaction does exhibit 
significant sensitivity to the type of $V_{\rm opt}$, it cannot yet 
be used to exclude potentials which are permissible from the point of view 
of fitting $K^-$ atomic data.

\section{Production of $K^-$ nuclear bound states} 
\label{sec:DD}

$K^-$ {\it nuclear} bound states are expected, generally, to have widths 
of order $\Gamma~\approx~-~2~{\rm Im}V_{\rm opt}(\rho_0) \approx 100$ 
MeV. However, for particularly deep states, bound by over 100 MeV, 
the dominant two-body pionic decay modes of the $\bar K N$ system 
get blocked so that these states stand a chance of being resolved 
in some appropriate production reactions. If such relatively narrow 
deeply bound states are ever observed, then the $K^-$ nucleus optical 
potential must be sufficiently deep. 
Kishimoto \cite{Kis99} has suggested to search for deeply 
bound $K^-$ nuclear states using the forward ($K^-,p$) reaction in which 
the incoming $K^-$ meson knocks out in the forward direction a bound proton 
from the target nucleus, while itself getting captured in a nuclear bound state 
generated by $V_{\rm opt}$. This is equivalent to a backward ($\theta = \pi$) 
$K^-p$ elastic scattering, and will be denoted by $K^-p \rightarrow pK^-$. 
In this section we calculate, more rigorously than done in Ref. \cite{Kis99}, 
the relevant production cross sections at laboratory momentum 
$p_{\rm L} = 1$ GeV/c. 

The forward differential ($K^-,p$) laboratory cross section on a nuclear target, 
due to a single-particle transition from the proton ($n_p,l_p$) shell to 
the $K^-$ ($n_{K^-},l_{K^-}$) bound state in the residual nucleus, is expressed 
in the distorted-wave (DW) impulse approximation \cite{HLW74} in terms of the 
Fermi-averaged forward $K^-p \rightarrow pK^-$ laboratory cross section: 
\beq 
\left(\frac{d\sigma (0^{\circ})}{d\Omega_{\rm L}} \right)^{(K^-,p)}\,= \, 
 \alpha \left(\frac{d\sigma (0^{\circ})}{d\Omega_{\rm L}} 
 \right)^{K^-p \rightarrow pK^-} P^{\rm DW}_{n_pl_p \rightarrow n_{K^-}l_{K^-}} \; .
\label{eq:xsec} 
\eeq 
Here, $\alpha$ is a kinematical factor which accounts for the transformation 
from the two-body laboratory system to the many-body laboratory system \cite{DLW80} 
and $P^{\rm DW}_{n_pl_p \rightarrow n_{K^-}l_{K^-}}$ is an 
effective proton number for the transition $n_pl_p \rightarrow n_{K^-}l_{K^-}$ 
given by 
\beq 
P^{\rm DW}_{n_pl_p \rightarrow n_{K^-}l_{K^-}}=\,\frac{1}{(2l_p+1)}\,
 \sum_{m_{K^-}m_p}\sum_{j_p = l_p \pm 1/2}S_{j_p}\, 
 \mid \langle (nlm)_{K^-}\mid {\chi^{(-)}_p}^{*}(\bx{r})\,
 \chi^{(+)}_{K^-}(\bx{r})\mid (nlm)_p {\rangle_{j_p}} \mid^{2} \; .
\label{eq:Peff} 
\eeq 
In Eq. (\ref{eq:Peff}), $S_{j_p}$ is the $j_p$ proton pickup 
spectroscopic factor in the target (with a maximum value of ($2j_p+1$)) 
and the suffix $j_p$ attached to the matrix element stands for a possible 
$j_p$ dependence of the bound proton radial wavefunction. We note that 
Eqs. (\ref{eq:xsec},\ref{eq:Peff}) are equivalent to 
Eq. (3.15) of Ref. \cite{ABD83} for the ($K^-,\pi^-$) reaction. 
For the distorted waves $\chi (\bx{r})$ we use the eikonal approximation, 
retaining only the exponential attenuation factor, 
\beq 
{\chi^{(-)}_p}^{*}(\bx{r})\,\chi^{(+)}_{K^-}(\bx{r}) \approx \exp (iqz) \, 
 \exp \left(-\frac{\bar \sigma}{2}T(b) \right) \; , 
\label{eq:eik} 
\eeq 
where $q$ is the momentum transfer, purely longitudinal at 0$^\circ$, with 
$\bar \sigma$ denoting an average $\bar {K} N$ and $pN$ total cross section, 
and where the nuclear thickness function $T(b)$ is defined by 
\beq 
T(b) = \int_{-\infty}^{\infty} \rho (r)\,dz  \;\;\;\;\;\;\;\;\; 
 (\int T(b)\, d^{2}b = A) \; . 
\label{eq:thick} 
\eeq 
Here $\bx{b}$ is the impact-parameter coordinate in the plane perpendicular 
to the direction of the forward momentum transfer $\bx{q}$.
The functions $T(b)$ were evaluated numerically, using realistic density 
distributions for $\rho (r)$.

We have calculated the DW expression (\ref{eq:Peff}) for the forward ($K^-,p$) 
reaction at incoming momentum $p_{\rm L}=1$ GeV/c on $^{12}$C and $^{28}$Si, 
to the $K^-$ $1s$ state generated by the $K^-$ nucleus DD optical potential. 
The imaginary part of $V_{\rm opt}$ was reduced to $12\%$ of its nominal 
strength in order to account for the reduced phase space for $K^-$ absorption, 
assuming no change in the in-medium properties of the decay products. The results 
given below are rather insensitive to the precise amount of this reduction. By 
approximating the $K^-$ and proton bound-state wavefunctions 
by harmonic oscillator (HO) wavefunctions, the matrix elements in Eq. 
(\ref{eq:Peff}) reduce to a one dimensional numerical integration in the 
variable $b$, as follows: 
\beq 
P^{\rm DW}_{1p_p \rightarrow 1s_{K^-}}\,=\,\left(\frac{{\tilde b}^8}
 {b_{K^-}^{3}b_{p}^5}\right)\,S_{p_{3/2}}\,\frac{1}{6}({\tilde b}q)^{2}\,
 \exp \left(-\frac{1}{2}({\tilde b}q)^{2}\right)\,{\mid G_{0}(\bar \sigma)\mid}^2 
\label{eq:PC}
\eeq
for $^{12}$C, and 
\beq 
P^{\rm DW}_{1d_p \rightarrow 1s_{K^-}}\,=\,\left(\frac{{\tilde b}^{10}}
 {b_{K^-}^{3}b_{p}^7}\right)\,S_{d_{5/2}}\,\frac{1}{15}\,\exp \left(-\frac{1}{2}
 ({\tilde b}q)^{2}\right)\,{\mid G_{2}(\bar \sigma)-G_{0}(\bar \sigma)
 (1-({\tilde b}q)^{2}/2)\mid}^2 
\label{eq:PSi} 
\eeq
for $^{28}$Si, 
with the distorted-wave integrals 
\beq 
G_{k}(\bar \sigma)\,=\,\int_{0}^{\infty} 2t^{k+1}\,\exp (-t^{2})\,
 \exp \left(-\frac{\bar \sigma}{2}T({\tilde b}t)\right)\,dt\;,\;\;\;\;
 G_{k}(\bar \sigma = 0)=1\;,\;\;\; k=0,2\;. 
\label{eq:Gs} 
\eeq 
The mean HO size parameter $\tilde b$ is defined by 
\beq 
\frac{1}{{\tilde b}^2}\,=\,\frac{1}{2}\left (\frac{1}{b_{K^-}^2}+\frac{1}{b_{p}^2} 
 \right ) 
\label{eq:bs} 
\eeq 
in terms of the $K^-$ and proton HO size parameters. 

The accuracy of this approximation is estimated to incur errors of up to $10\%$ 
in the calculated cross sections. In order to provide a concrete and useful 
check, we have also calculated the forward ($\pi^+,K^+$) reaction 
cross sections at $p_{\rm L}=1.04$ GeV/c 
on the same targets as above, leading to the known $1s$ $\Lambda$ hypernuclear 
ground states in the residual nuclei. The momentum transfer characterizing this 
reaction is quite similar to that in the ($K^-,p$) reaction. The calculation of 
cross sections for the production of $\Lambda$ $1s$ bound states is identically 
the same as for the $K^-$ $1s$ bound states, upon the replacement of bound 
protons in Eqs. (\ref{eq:xsec},\ref{eq:Peff}) by bound neutrons. The results of 
this calculation are shown in Table \ref{tab:piK}, with $b_{\Lambda}$ and $b_n$ 
standing for the $\Lambda$ and neutron HO size parameters, respectively, 
where the size parameter $b$ is identified by the $\exp (-r^2/2b^2)$ exponential 
factor of the HO wavefunction. The ratio of effective neutron number 
for the DW calculation, $N^{\rm DW}$, to the effective neutron number for the 
plane-wave (PW) calculation, $N^{\rm PW}$, is also given, as a measure of the 
effect of absorption. The input parameters 
$\alpha~(d\sigma (0^{\circ})/d\Omega_{\rm L}) = 0.5$ mb/sr for 
$\pi^{+}n \rightarrow K^{+}\Lambda$, $\bar {\sigma} = 27.5$ mb and 
$S_{j_n}=2j_{n}+1$ for the valent $p_{3/2}$ and $d_{5/2}$ neutron orbits 
in $^{12}$C and $^{28}$Si, respectively, are the same as in the eikonal 
calculation of Motoba et al. \cite{MBW88}, which did not use the HO 
approximation, for the ($\pi^+,K^+$) reaction. 
The agreement between the latter calculation and the present one, 
as shown in Table \ref{tab:piK}, is very reasonable, inspiring confidence 
in the present results for the ($K^-,p$) reaction (Table \ref{tab:Kp}). 
We note that the preliminary value for the ($\pi^+,K^+$) forward cross 
section on $^{12}$C in experiment E336 at KEK \cite{Has00} is $15 \pm 1$ 
$\mu$b/sr. 

The forward ($K^-,p$) differential cross sections listed in Table \ref{tab:Kp} 
were calculated assuming a value of 3.6 mb/sr for the factor 
$\alpha~(d\sigma (0^{\circ})/d\Omega_{\rm L})$ in front of $P^{\rm DW}$ on the 
r.h.s. of Eq. (\ref{eq:xsec}). This value is based on the measured two-body 
$K^-p$ c.m. backward cross section $1.7 \pm 0.1$ mb/sr at $p_{\rm L}$ = 1 GeV/c 
\cite{Con76}, on the value $\alpha = 0.69$ appropriate to this kinematics, 
and on an estimated Fermi-average reduction factor of 0.59 due to the peaking 
of the two-body elastic $K^-p$ backward cross section at 1 GeV/c as function 
of the incoming momentum, following the procedure outlined in Ref. \cite{ABD83}. 
For the distortion, using the value $\bar \sigma$ = 40 mb \cite{Kis99}, 
the suppressive effect on the forward ($K^-,p$) cross sections is considerably 
stronger than for the ($\pi^+,K^+$) reaction. This is partly due also to the 
extremely small spatial extension of the $K^-$ wavefunction which restricts the 
($K^-,p$) reaction to the denser nuclear region where absorption prevails. 
In contrast, the extension of the $\Lambda$ wavefunction in the ($\pi^+,K^+$) 
reaction is substantially larger. The effect of absorption gets stronger with 
the atomic number of the target nucleus, as seen clearly from 
Table \ref{tab:Kp} by comparing the results for Si with those for C. 
Another reason for the fast decrease of the calculated cross section with 
increasing $A$ is the gradual increase of the momentum transfer $q$ in the 
($K^-,p$) reaction due to the increased binding of the $K^-$ meson. This 
is just opposite to the trend observed in Table \ref{tab:piK} for the 
($\pi^+,K^+$) reaction, where $q$ gradually decreases as function of $A$ 
due to the increased binding of the $\Lambda$ hyperon. 

Finally, we comment on the marked disagreement between the results of the 
present calculations and the estimates due to Kishimoto \cite{Kis99} shown 
in Table \ref{tab:Kp}. The wide range of estimated cross-section values 
in Ref. \cite{Kis99} reflects primarily the dependence of the PW expressions 
Eqs. (\ref{eq:PC},\ref{eq:PSi}) on the HO size parameter $b_{K^-}$. 
The smaller $b_{K^-}$ is, the higher the PW cross section becomes. 
However, the lowest value of $b_{K^-}$ assumed there, $b_{K^-}=b_N/8^{1/4}$, 
was erroneously chosen instead of the considerably larger value 
$b_{K^-}=b_N/2^{1/4}$ that the correct HO scaling argument leads to. 
This should narrow appreciably the range of values for the PW cross section. 
Furthermore, it appears inconceivable that just one value for the 
distortion factor $P^{\rm DW}/P^{\rm PW}$, as quoted from Ref. \cite{Kis99} 
in Table \ref{tab:Kp}, 
can be considered representative for the whole range of $b_{K^-}$ values 
assumed by Kishimoto. For this matter, the smaller $b_{K^-}$ is, the 
more suppressive is the effect of the distortion. The dependence on 
$b_{K^-}$ is rather strong and thus largely cancels the 
opposite trend of the PW calculation as explained above. The overall 
dependence on $b_{K^-}$ is therefore much weaker than one is led to 
believe in Ref. \cite{Kis99}. As for the absolute scale of the suppression 
provided by the distortion factors $P^{\rm DW}/P^{\rm PW}$, we find it 
inconceivable that Kishimoto's suppression effects are even weaker 
than we and other works \cite{MBW88} find for the ($\pi^+,K^+$) reaction. 
The present results call for extreme caution when contemplating ($K^-,p$) 
experiments aimed at identifying deeply bound $K^-$ nuclear states.

\section{Summary and Conclusions}
\label{sec:end}

At present the best tool for exploring the $K^-$ interactions in the nuclear 
medium at low energy is the study of strong interaction effects in kaonic 
atoms. Extrapolating this interaction to higher densities, as encountered 
in astrophysical scenarios, must rely on some theory, beyond the 
phenomenological potential that fits the data very well. The prime aim 
of the present work was to see how far one can go in bringing microscopic 
approaches to the $K^-$-nucleus interaction into agreement with the $K^-$ 
atomic data. The chirally motivated coupled channel approach to the 
${\bar K}N$ interaction, which is quite successful in reproducing all the 
low energy $K^-p$ data, was chosen as a starting point. Earlier attempts 
to use this approach indicated poor agreement with the atomic data and only 
by introducing self consistency into the theory it became possible to achieve 
barely acceptable fits. Empirical modifications of the interaction managed to 
improve the fits to the atomic data, but at the cost of losing contact 
with the underlying $\bar KN$ interaction model, thus making questionable 
any extrapolation of such optical potentials to higher densities.

In the present work we addressed the problem by requiring {\it simultaneous} 
fits to atomic and $K^-p$ data within the chirally motivated coupled 
channel approach, including self consistently the self energies of the 
kaons and, in some cases, of the nucleons. 
It was found that minor modifications of parameters of this theory led 
to good agreement with the atomic data while maintaining the good agreement 
with the $K^-p$ data. In fact, we found that one can fit only the atomic 
data and still not lose the good fit to the $K^-p$ data. The best 
value of $\chi^2$ per point obtained here for the $K^-$ atomic data is 2.2, 
compared to 1.5 for the best-fit phenomenological potential. 
Nevertheless, the fits of the self consistent chirally motivated 
coupled channel potentials are quite good. 

The depth of the real part of the $K^-$ nucleus potential is also of 
prime interest in connection with the possibility of kaon condensation 
in collapsing stars. Deep (180 MeV) potentials were found in the 
phenomenological analysis, whereas shallow (55 MeV) potentials are 
found in the framework of models which require the $K^-$ optical potential 
to be derived self consistently. It seems impossible to reconcile the 
phenomenological potentials with the present microscopic potentials 
both in terms of depth and in terms of values of $\chi^2$. Since the 
depth of the $K^-$-nucleus optical potential cannot be resolved by 
studying only kaonic atoms, we briefly discussed the ability to do so 
with the help of reactions initiated by stopped $K^-$~mesons. 
The mechanism behind it is the sensitivity of the DW integrals to
the depth of the real part of the $K^-$ potential, due to the node
structure of the wavefunction, which is quite different from each other 
for deep and for shallow potentials. It was demonstrated that 
$\Lambda$-hypernuclear ground-state production rates 
calculated for the ($K^-_{\rm stop},\pi$) reactions on carbon 
differ by more than a factor of 3 between the different potential 
depths mentioned above. Unfortunately, these calculated rates are 
still several times smaller than the measured ones \cite{THO94}, 
so that these reactions cannot yet be used to reach a definite 
conclusion on the depth of the $K^-$ nucleus optical potential near
threshold. However, it is plausible that by studying the effect 
of the $K^-$ potential on the pion spectrum of the capture reactions, 
one could reach more definite conclusions.

Finally, we addressed the possibility of identifying deeply bound
$K^-$ {\it nuclear} states in the forward ($K^-,p$) reaction on 
nuclear targets. Using the very deep DD potential \cite{FGB94}, 
we calculated a cross section as large as 47 $\mu$b/sr at 
$p_{\rm L}=1$ GeV/c for the production of the $1s$ $K^-$ bound 
state on $^{12}$C. This $K^-$ nuclear `ground state' is bound 
by about 120 MeV, which strips it off most of the phase space 
for its dominant pionic decay modes, so that its residual width 
could become as small as 10-20 MeV. In contrast, using the HO 
size parameter $b_{K^-}=1.22$ fm from Table \ref{tab:Kp}, 
the first excited ($1p$) state is estimated to lie about 
$\hbar \omega \approx 50$~MeV higher and, therefore, it should 
be much wider and hardly observable. We verified this estimate by an
explicit calculation. If the $K^-$ nuclear potential is not of the 
`deep' kind, it may still accommodate bound states, but these will be 
very wide, judging from the depth of Im $V_{\rm opt}$ found in this 
work for the `shallow' chirally motivated potentials. We calculated 
the $1s_{K^-}$ bound state production cross section also on $^{28}$Si, 
finding it to be about 8 times smaller than on $^{12}$C. This rules out 
using medium-weight or heavy targets for this reaction. Not much 
different production cross sections should be expected at lower 
incoming momentum, say at 600 MeV/c. Our calculated cross sections, 
tested for the kinematically similar ($\pi^+,K^+$) reaction, are 
substantially lower than those estimated by Kishimoto \cite{Kis99}.

\vspace{8mm}
This research was partially supported by the Israel Science Foundation 
(E.F. and A.G., grant No. 171/98) and by the Grant Agency of the Czech 
Republic (A.C. and J.M., grant No. 202/00/1667). 
A.C. and J.M. acknowledge the hospitality of the Hebrew University. 
We thank A. Ramos and J. Schaffner-Bielich for discussions and for 
communicating the scattering amplitudes of Refs. \cite{ROs00,SKE00} 
respectively.

\begin{table}
\caption{Kaonic atom optical potentials using two sets of self consistent 
in-medium $\bar K N$ scattering amplitudes,  
with or without adding a complex `$t\rho$' potential, and with or without 
scaling the $I=1$ amplitude. $V_{\rm R}$ and $V_{\rm I}$ refer to 
the depth of $V_{\rm opt}$ for Ni.}
\label{tab:ROS}
\begin{tabular}{lccccccc}
amplitude & $t_{\rm R}$ (fm)&$t_{\rm I}$(fm)&$s_{\rm R}$&$s_{\rm I}$&$
\chi ^2/N$&$V_{\rm R}$(MeV)
&$V_{\rm I}$(MeV)  \\
\hline
Ref.\cite{ROs00}&0&0&1&1&4.62&$-$44&$-$54 \\
 &0.11$\pm$0.02&$-$0.20$\pm$0.02&1&1&2.73&$-$58&$-$23 \\
 &0&0&2.0$\pm$0.2&0.9$\pm$0.2&3.08&$-$61&$-$52 \\
 & 1.17$\pm$0.11&$-$0.03$\pm$0.15&$-$0.5$\pm$0.5&0.96$\pm$0.54&
2.00&$-$170&$-$49 \\
\hline
Ref.\cite{SKE00}&0&0&1&1&12.7&$-$34&$-$62 \\
 & 1.0$\pm$0.1&0.26$\pm$0.06&1&1&2.46&$-$159&$-$95 \\
 &0&0&4.5$\pm$0.3&5.7$\pm$0.7&2.82&$-$134&$-$116 \\
 &2.9$\pm$0.4&$-$2.8$\pm$0.8&$-$6.1$\pm$1.7&34$\pm$9&1.49&$-$204& $-$82 \\
\end{tabular}
\end{table}

\begin{table}
\caption{Calculated branching ratios (\ref{eq:ratios}) at the $K^{-}p$ 
threshold, and total cross sections (in mb) for selected $K^{-}p$ reactions 
(\ref{eq:xsecs}) at incident momentum $p_{\rm L}=110$~MeV/c, for the `no SC' 
and `$V_{\rm N}$, full' parameterizations of Table \ref{tab:coupl}. 
Also shown are the experimental data.}
\label{tab:ratios}
\begin{tabular}{cccccccc}
 & $\gamma$ & $R_{\rm c}$ & $R_{\rm n}$ & $\sigma(K^{-}p)$ & $\sigma(\bar{K}^{0}n)$
    & $\sigma(\pi^{+}\Sigma^{-})$ & $\sigma(\pi^{-}\Sigma^{+})$ \\ \hline
`no SC'  & 2.383 & 0.667 & 0.171 & 93.2 & 41.2 & 60.1 & 32.3 \\
`V$_{\rm N}$, full' & 2.347 & 0.669 & 0.198 & 94.5 & 38.0 & 71.2 & 42.0 \\ \hline
exp.\cite{Mar81,Cib82,Eva83} & 2.36$\pm$0.04 & 0.664$\pm0.011$ & 0.189$\pm$0.015
    & 92 $\pm$ 8 & 29 $\pm$ 6 & 64 $\pm$ 10 & 29 $\pm$ 6 \\
\end{tabular}
\end{table}

\begin{table}
\caption{Values of $\chi ^2$, optical potentials and scaling factors for 
the present chiral model. $\chi ^2_{\rm atom}$ refer to the $K^-$ atom 
65 data points and $\chi ^2_{\rm T}$ is the sum of the former and the 
$\chi ^2$ for the 7 data points for the free-space 
$K^-p$ data. $V_{{\rm R,I}}$ are the optical potential values 
at the center of the Ni nucleus and $V_{{\rm N}}$ stands 
for fits where the nucleon optical potential is included. 
The scaling factors $f$ and $\alpha (j)$ multiply the 
parameters $f_{\pi}$ and $\alpha_j$, respectively, 
of the coupled channel model.}
\label{tab:coupl}
\begin{tabular}{lccccccccc}
model & $\chi ^2_{{\rm T}}$ & $\chi ^2_{{\rm atom}}$ & $V_{{\rm R}}$(MeV) &
  $V_{{\rm I}}$(MeV) & $f$ & $\alpha(K^-p)$ &
      $\alpha(\bar K^0n)$ & $\alpha(\pi^0 \Lambda)$ &
            $\alpha(\pi\Sigma)$ \\ \hline
no SC &1082 & 1076 & $-$117 & $-$67 & 1 & 1 & 1 & 1 & 1 \\
 SC & 443.1 & 436.7 & $-$70.6 & $-$85.6 & 1 & 1 & 1 & 1 & 1 \\
atoms & 169.9 & 157.2 & $-$66.9 & $-$55.2& 1.12$\pm$0.05 & 1.00$\pm$0.08 &
     1.21$\pm$0.08 & 1.05$\pm$0.17 & 1.39$\pm$0.07 \\
full & 165.3 & 157.5 & $-$66.8 & $-$54.0 & 1.14$\pm$0.05 & 1.03$\pm$0.07 &
     1.24$\pm$0.08 & 1.13$\pm$0.16 & 1.35$\pm$0.07 \\
$V_{{\rm N}}$, atoms & 156.3 & 145.4 & $-$56.3 & $-$61.8 & 1.04$\pm$0.04 &
    0.86$\pm$0.06 & 1.07$\pm$0.05 & 1.00$\pm$0.14 & 1.18$\pm$0.08 \\
$V_{{\rm N}}$, full & 154.0 & 145.8 & $-$54.9 & $-$60.2 & 1.06$\pm$0.04 &
    0.90$\pm$0.06 & 1.10$\pm$0.06 & 1.08$\pm$0.15 & 1.17$\pm$0.07 \\
\end{tabular}
\end{table}

\begin{table}
\caption{$K^-$ optical potentials and calculated strong-interaction 
shifts and widths, in eV, for the $2p$ and $3d$ $K^-$ atomic levels 
in $^{12}$C.}
\label{tab:C1}
\begin{tabular}{cccccccc}
$V_{\rm opt}$ & $a$ (fm) & $B$ (fm) & $\gamma$ & $\epsilon_{2p}$
  & $\Gamma_{2p}$   & $\epsilon_{3d}$
  & $\Gamma_{3d}$  \\ \hline 
\mbox{chiral} & - & - & - & $-616$ & $1484$ & $+0.01$ & $0.57$ \\
\mbox{$t_{\rm eff}$} & $ 0.63+{\rm i}0.89$ &   -    &  -
      & $-590$ & $1360$ & $-0.01$ & $0.64$ \\
\mbox{$\tilde{t}_{\rm eff}$}  & $1.30+{\rm i}0.80$  &   -    &  -
      & $-605$ & $1731$ & $-0.01$ & $0.85$ \\ 
\mbox{DD} & $-0.15+{\rm i}0.62$ & $1.65-{\rm i}0.06$ & $0.23$
      & $-468$ & $1578$ & $-0.08$ & $0.64$ \\ \hline

exp.\ \cite{BAC72} & & & & $-590 \pm 80$ & $1730 \pm 150$ & - & 
$0.98 \pm 19$ \\ 
\end{tabular}
\end{table}

\begin{table}
\caption{Calculated capture rates on $^{12}$C per stopped $K^-$ 
(in units of $10^{-3}$) to the summed $p_{N} \rightarrow s_{\Lambda}$ 
$1^{-}$ excitations in $^{12}_{\Lambda}$C and $^{12}_{\Lambda}$B, 
for the $K^-$ optical potentials of Table \ref{tab:C1}.} 
\label{tab:C2}
\begin{tabular}{ccccc}
final $_{\Lambda}^{\rm A}\rm Z$ & \mbox{chiral} & \mbox{$t_{\rm eff}$} 
 & \mbox{$\tilde{t}_{\rm eff}$} & \mbox{DD} \\ \hline 
$_{\Lambda}^{12}\rm C$ & $0.231$ & $0.169$ & $0.089$ & $0.063$ \\ 
$_{\Lambda}^{12}\rm B$ & $0.119$ & $0.087$ & $0.046$ & $0.032$ \\ 
\end{tabular}
\end{table}

\begin{table}
\caption{Input and results of DWIA eikonal calculations for the forward
$(\pi^{+},K^{+})$ reaction cross section (in $\mu$b/sr) at incoming
momentum $p_{\rm L}$ = 1.04 GeV/c, see text for details.}
\label{tab:piK}
\begin{tabular}{lccccccc}
	target & $B^{\Lambda}_{1s}$& $q$ & $b_{\Lambda}$& $b_{n}$&
	$N^{\rm DW}/N^{\rm PW}$ &
     \multicolumn{2}{c}{$({{d\sigma(0^{\circ})}/{d
	\Omega_{\rm L}}})^{(\pi^{+},K^{+})}$}   \\
	nucleus   &(MeV) &(MeV/c) & (fm) & (fm) &[present]&[present]&
        Ref. \cite{MBW88}  \\
	\hline
	$^{12}$C  & 10.8 & 335    & 1.72 & 1.52 & 0.237   & 15.4    &
        17.4          \\
	$^{28}$Si & 16.7 & 322    & 1.92 & 1.72 & 0.132   &  7.7    &
         8.9          \\
\end{tabular}
\end{table}

\begin{table}
\caption{Input and results of DWIA eikonal calculations for the forward
$(K^{-},p)$ reaction cross section (in $\mu$b/sr) at incoming momentum
$p_{\rm L}$ = 1 GeV/c, see text for details.}
\label{tab:Kp}
\begin{tabular}{lccccccccc}
	target & $B^{K^{-}}_{1s}$& $q$ & $b_{K^{-}}$& $G_{0}$ & $G_{2}$ &
    \multicolumn{2}{c}{	$P^{\rm DW}/P^{\rm PW}$} &
   \multicolumn{2}{c}{$({{d\sigma(0^{\circ})}/{d
	\Omega_{\rm L}}})^{(K^{-},p)}$}    \\
	nucleus   &(MeV)&(MeV/c)& (fm) &       &       & [present] &
        Ref. \cite{Kis99} & [present] & Ref. \cite{Kis99}  \\
	\hline
	$^{12}$C  & 122 &  369  & 1.22 & 0.308 &  ---  & 0.095     &
        0.25         & 47       & 100-490        \\
	$^{28}$Si & 144 &  404  & 1.42 & 0.183 & 0.269 & 0.040     &
        0.16         &  6.0      &  35-180         \\
\end{tabular}
\end{table}

\end {document}